\documentclass[aps,pra,floatfix,showpacs,twocolumn,superscriptaddress]{revtex4}

\usepackage{amsmath}
\usepackage{graphicx}

\newcommand{\bra}[1]{\mbox{$\langle #1|$}}
\newcommand{\ket}[1]{\mbox{$|#1\rangle$}}

\newcommand{\Ket}[2]{|#1\rangle^{(#2)}}

\newcommand{\XX}{X\!X\!}
\newcommand{\XXp}{X\!X'}
\newcommand{\fI}{f_\mathrm{I}}
\newcommand{\fII}{f_\mathrm{II}}

\begin{document}

\title{Fault Tolerance in Parity-State Linear Optical Quantum Computing}

\author{A.~J.~F.~Hayes}
\email{ahayes@physics.uq.edu.au}
\affiliation{Centre for Quantum
Computer Technology and Physics Department, University of Queensland, QLD 4072, Brisbane,
Australia.}

\author{H.~L.~Haselgrove}
\affiliation{C3I Division, Defence Science and Technology Organisation,
Canberra, 2600, Australia.}

\author{Alexei~Gilchrist}
\affiliation{Physics Department, Macquarie University, Sydney, NSW 2109, Australia.}

\author{T.~C.~Ralph}
\affiliation{Centre for Quantum
Computer Technology and Physics Department, University of Queensland, QLD 4072, Brisbane,
Australia.}

\date{\today}

\begin{abstract}
We use a combination of analytical and numerical techniques to calculate the noise threshold and resource requirements for a linear optical quantum computing scheme based on parity-state encoding. Parity-state encoding is used at the lowest level of code concatenation in order to efficiently correct errors arising from the inherent nondeterminism of two-qubit linear-optical gates. When combined with teleported error-correction  (using either a Steane or Golay code) at higher levels of concatenation, the parity-state scheme is found to achieve a saving of approximately three orders of magnitude in resources when compared to a previous scheme, at a cost of a somewhat reduced noise threshold.
\end{abstract}


\maketitle
\section{Introduction}

It was shown by Knill, Laflame, and Milburn (KLM) \cite{klm} that,
in principle, scalable optical quantum computing could be achieved using only passive linear
elements, single-photon sources, measurement and feedforward. Non-deterministic gates were developed that failed through the accidental measurement of qubit value. Parity state error codes were developed to protect against such accidental measurement. KLM showed that, by concatenating the parity codes, gate failures could be reduced to arbitrarily small levels, thus justifying the claim of scalibility. However, because of the massive complexity of the scheme, it has not been practical to couple it to a higher level error correction protocol capable of correcting environmental errors, and hence evaluate its resource requirements and fault tolerant threshold.

A major simplification of the KLM circuit approach was achieved by the introduction of incremental parity codes \cite{hay04} and fusion gate techniques \cite{bro05,gil07}. We refer to this modification of KLM as parity state quantum computing. These techniques reduce the complexity of the scheme sufficiently that it becomes possible to make a full fault tolerant analysis, thus completing the original KLM program.

In this paper, we derive the resource usage and error thresholds achievable when a concatenated error-correcting code such as the Steane code \cite{ste96} is used to to handle environmental and residual gate errors in the parity-state optical quantum computing scheme. This type of analysis has previously been done for cluster state \cite{daw06} and coherent state \cite{lund08} schemes. It is important to establish these thresholds for the parity-state implementation both as a target for technological development, and for comparison with the other proposals. Our results show that the parity-state protocol may offer a useful trade-off between the higher resource usage of cluster-state schemes and the lower noise threshold of coherent state schemes.

Significant progress has been made in optical quantum computing experiments in the last decade \cite{kok07}.
In particular the basic principles of optical parity state production and their ability to correct $Z$-measurement errors has been demonstrated \cite{pitt05,obr05,lu08}. These promising experiments indicate that effective use of such error-correcting codes in future designs is viable.

The layout of the paper as follows.  Section \ref{secgates} gives a brief review and introduction of parity encoding and the operations that may be done on parity-encoded states. Section \ref{secerrmod} describes the physical noise model that we consider, and gives expressions for the effective noise rates on various parity-encoded operations. In Section \ref{secthresh} these error-rate expressions are used as the basis of simulations of higher levels of encoding. The results of the simulations are presented in the form of noise-threshold curves. Finally, Section \ref{secres} considers the resources required by this scheme, and provides a comparison with the thresholds
and resource requirements of some other schemes for fault-tolerant optical quantum computation.

\section{Universal gate set}\label{secgates}

This section describes the states and operations used in the two lowest levels of encoding in our scheme: the physical qubits in the dual-rail nondeterministic linear-optical architecture, and the first level of logical qubits which use parity-state encoding. These designs lead to the central focus of this paper, a discussion of the effects of noise on the parity-state encoding and the effects of higher levels of encoding (fault-tolerant Steane and Golay encoding), which is covered in Sections~\ref{secerrmod}--\ref{secres}.

\subsection{Physical Encoding and Operations}


At the physical level, qubits in our scheme are encoded and manipulated according to the techniques of nondeterministic dual-rail linear optics. Although other implementations, such as spatial encoding, are possible, we will explicitly consider polarisation qubits, encoded in the horizontal and vertical polarisation modes of single photons. This entails a series of physical and technological assumptions including: that single-photon states can be produced on-demand in a desired mode,  such modes can be made to interact using mode-matched linear optics, modes can be stored in a quantum memory, and, the number of photons in a mode can be measured and the results used in the fast control of optical switching. Single-qubit operations in this scheme are relatively straightforward, but two-qubit gates are inherently nondeterministic even in the absence of noise.

\begin{figure}
\begin{center}
\includegraphics[width=5.0cm]{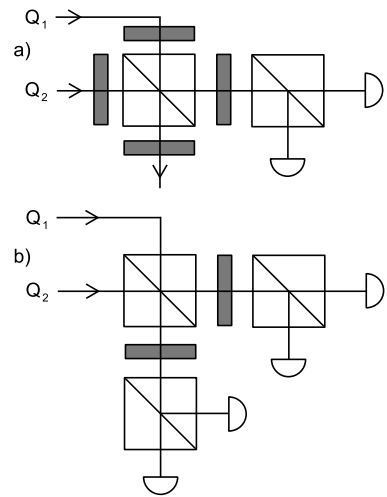}
\caption{a) The type-I fusion gate. b) The type-II fusion gate. Here the gates are shown being performed on two polarization-encoded photonic input qubits, $Q_1$ and $Q_2$. These input qubits are typically part of larger entangled states. The gates are constructed from photon-counting detectors, polarising beam-splitters, and half-wave plates, where the half-wave plates act as Hadamard operations on the polarization qubits.} \label{fusiongates}
\end{center}
\end{figure}

As in reference \cite{bro05}, our scheme utilises two particularly simple nondeterministic two-qubit gates, the so-called
type-I ($\fI$) and
type-II ($\fII$) fusion gates (Fig~\ref{fusiongates}). A type-II fusion gate performs a two-qubit destructive measurement in the basis $\{\ket{00}+\ket{11},\ket{00}-\ket{11},  \ket{01},\ket{10}\}$. The first two outcomes (corresponding to maximally-entangled basis elements) are considered ``successful'', whereas the second two outcomes are considered to be failures of the gate. A type-I fusion gate is a partial Bell measurement on two qubits. Two outcomes, considered successful, project the input state into the space spanned by $\{\ket{00},\ket{11}\}$ and outputs a single qubit according to the operator $\ket{0}\bra{00} \pm \ket{1}\bra{11}$. Again, there are two failure outcomes which correspond to a destructive measurement in the basis $\{\ket{01},\ket{10}\}$.

For the type of input states we will consider, both the type-I and type-II fusion gates are successful with 50\% probability (in the absence of noise).

\subsection{Parity Encoding}
A length-$n$ {\em parity code} encodes one logical qubit into $n$ physical qubits. The logical basis states of the code, denoted $\Ket{0}{n}$ and $\Ket{1}{n}$, are defined to be:
\begin{eqnarray}
\label{parity}
\Ket{0}{n} & \equiv & (\ket{+}^{\otimes n}+\ket{-}^{\otimes
n})/\sqrt{2}\nonumber \\
\Ket{1}{n} & \equiv & (\ket{+}^{\otimes n}-\ket{-}^{\otimes
n})/\sqrt{2},
\end{eqnarray}
where  $\ket{\pm} = (\ket{0} \pm \ket{1})/\sqrt{2}$. Note that $\Ket{0}{n}$ is the equal superposition of all even-parity $n$-bit strings, and  $\Ket{1}{n}$ is the equal superposition of all odd-parity strings.
A useful property of the parity code basis states is that they have a simple expansion in terms of smaller code states:
\begin{eqnarray}
\label{parityexpand}
\Ket{0}{n} & = & (\Ket{0}{n-j}\Ket{0}{j}+\Ket{1}{n-j}\Ket{1}{j})/\sqrt{2}, \\
\Ket{1}{n} & = & (\Ket{1}{n-j}\Ket{0}{j}+\Ket{0}{n-j}\Ket{1}{j})/\sqrt{2},
\end{eqnarray}
where $1 \le j \le n-1$. For $j=1$ this expansion shows that a computational
basis measurement (such as occurs with a failed fusion gate) of one of the physical qubits will not destroy
a parity-encoded qubit, but will only reduce the length of encoding by one (and possibly introduce a known logical Pauli $X$ operation, depending on the measurement outcome).

\subsection{Generating Parity States}

The production of parity-encoded states is a necessary procedure
both for the preparation of encoded qubits as sources, and as part
of the resource generation required by some of the non-deterministic
logical gates in our universal gate set for parity-encoded qubits.

The state $\Ket{0}{n}$ is locally equivalent to a star shaped cluster state (by
a Hadamard operation applied to the central node of the star).  Consequently,
given a supply of Bell states ($\Ket{0}{2}$), the resource $\Ket{0}{n}$ can be
built up using essentially the same techniques as used in \cite{bro05} to build
up star-shaped cluster states.

Parity states $\Ket{0}{n}$ and $\Ket{0}{m}$ can be {\em fused} using the $\fI$ gate as follows:
\begin{equation}
       H \fI(H\!\otimes\! H)   \Ket{0}{n}\Ket{0}{m} \rightarrow \left\{ \begin{array}{cr}
               \Ket{0}{m+n-1}  & \mbox{(success)}\\
               |+\rangle^{\otimes n-1} |-\rangle^{\otimes m-1} & \mbox{(failure)}
       \end{array}\right.\label{eq:fIjoin}
\end{equation}
(where we have omitted other possible locally equivalent outcomes for both success and failure). The operator  $H \fI(H\!\otimes\! H)$ should be understood as a Hadamard gate acting on one of the physical qubits from each of the encoded states, followed by the $\fI$ gate applied to the same pair, followed by a Hadamard gate applied to the output of $\fI$ in the case of success. Alternatively the $\fII$ gate can be used to carry out fusion, as follows:
\begin{equation}
       \fII \Ket{0}{n}\Ket{0}{m}\rightarrow \left\{ \begin{array}{cl}
               \Ket{0}{m+n-2} & \mbox{(success)}\\
               \Ket{0}{m-1}\Ket{0}{n-1} & \mbox{(failure)}
       \end{array}\right.
       \label{eq:fIIjoin}
\end{equation}
(again, some additional locally-equivalent outcomes have been omitted).

The first alternative, where $\fI$ is used with Hadamard gates to perform fusion, has the advantage of losing only a single physical qubit from the
input states, but the disadvantage of completely destroying the
entanglement in both input states in the event of failure. In the
second case, $\fII$ is used to join the input states at the expense
of losing two of the initial physical qubits. There are three
advantages to the second scheme --- in the case of failure we do not
destroy the entanglement of the input states, just reduce their
encoding by one; we do not need photon number discriminating
detectors to operate $\fII$ ; and it is failsafe with respect to
loss, in the sense that a lost photon will not cause a failed $\fII$ gate to appear to have operated successfully.

Thus, to create the state $\Ket{0}{3}$, two $\Ket{0}{2}$ are fused
together using the $\fI$ gate. Since $\fI$ functions with a probability of $1/2$, on
average two attempts are necessary so on average each $\Ket{0}{3}$
consumes $4 \Ket{0}{2}$. Once there is a supply of $\Ket{0}{3}$ states, either $\fI$ or
$\fII$ can be used to progressively build up larger resource states.

\subsection{Simple Single-Qubit Gates}
For the parity code, encoded single-qubit unitaries can be divided into those which have a particulary simple deterministic implementation, and those which have a more complicated nondeterministic implementation involving the consumption of resource states.

We can deterministically perform encoded versions of any of the
gates in the set $\{X_\theta, Z\}$. Here, $X_\theta$ refers to an arbitrary rotation about the $X$-axis of the Bloch sphere, $X_\theta = \cos(\theta/2)I+i\sin(\theta/2)X$.
An encoded  $X_\theta$  operation is achieved by applying $X_\theta$ to just one physical qubit in the code state. The encoded $Z$ gate is achieved by
applying a $Z$ gate transversally to all physical qubits in the code state.

\subsection{$Z_{90}$ Gate}

To make our set of encoded single-qubit gates universal, we add the $Z_{90}$ operation. Similar to the notation introduced in the previous subsection, $Z_{90}$ refers to a rotation by 90 degrees around the $Z$-axis of the Bloch sphere.

The logical $Z_{90}$ operation is based on the process of {\em re-encoding}.
Re-encoding can be understood by considering the following generalization of Equation~\ref{eq:fIIjoin}:
\begin{equation}
       \fII \Ket{\Psi}{n}\Ket{0}{m}\rightarrow \left\{ \begin{array}{cl}
               \Ket{\Psi}{m+n-2} & \mbox{(success)}\\
               \Ket{\Psi}{m-1}\Ket{0}{n-1} & \mbox{(failure)}
       \end{array}\right.
       \label{eq:fIIjoinB},
\end{equation}
where $\Ket{\Psi}{n}\equiv \alpha\Ket{0}{n} + \beta\Ket{1}{n}$ is an arbitrary parity-encoded input state, and like Equation~\ref{eq:fIIjoin} we have omitted other locally-equivalent outcomes.

To re-encode a logical qubit $\Ket{\Psi}{n}$, a type-II fusion gate is
first performed between the logical qubit and another ancillary parity state $\Ket{0}{n+1}$. Then, each of the remaining $n-1$ qubits that belonged to the original encoded input state are measured in the computational basis,
leaving the new ancilla qubits in the same state as the original input,  $\Ket{\Psi}{n}$. A logical $X$ operation
may be required as a correction depending on the total parity of the measurements made.

A slight modification of this procedure yields the encoded $Z_{90}$ gate.
Let
\begin{eqnarray} \Ket{\Psi}{n}=\alpha
(\Ket{0}{n-1}\ket{0}_I+\Ket{1}{n-1}\ket{1}_I)+\\ \beta
(\Ket{0}{n-1}\ket{1}_I+\Ket{1}{n-1}\ket{0}_I)\nonumber
\end{eqnarray}
be the logical qubit on which we wish to perform an encoded $Z_{90}$ operation. One
of the component physical qubits (here denoted by the subscript $I$)
is chosen to represent the input for the type-II fusion in the re-encoding procedure. To achieve an encoded $Z_{90}$ gate on $\Ket{\Psi}{n}$, we simply apply the un-encoded $Z_{90}$ gate to qubit $I$, then carry out the re-encoding procedure detailed above. The final logical state following a
successful fusion is $Z_{90}\ket{\Psi}$. In this case, a
correction corresponding to a logical $Y$ operation may need to be applied to the output state depending on the parity of the measurements made. In the event that the fusion
gate fails, the size of the input state is reduced, and the operation may
be re-attempted if there are enough remaining qubits. If all qubits in the input state are depleted, then the logical gate is considered to have failed.


To implement a logical Hadamard operation in this gate set we use the decomposition $H=X_{90}Z_{90}X_{90}$.
Since operations of the form $X_{\theta}$ are relatively easy to perform, the logical Hadamard is
essentially equivalent to the $Z_{90}$ gate in terms of difficulty, time taken, and error emergence.


\subsection{$\XXp_{90}$ Gate} \label{subsec:xx90}

\begin{figure}
\begin{center}
\includegraphics[width=6cm]{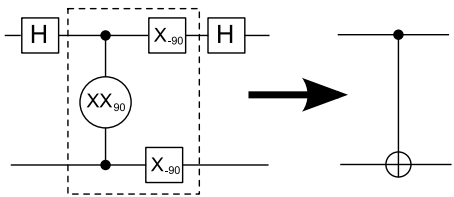}
\caption{The relationship between the $\XX_{90}$ gate and the {\sc cnot} gate. The combined operation inside the dotted area defines the $\XXp_{90}$ gate.} \label{xxcn}
\end{center}
\end{figure}

We define two maximally-entangling two-qubit gates $\XX_{90}$ and $\XXp_{90}$ as follows:
\begin{eqnarray}
\XX_{90}&\equiv&\frac{1}{\sqrt{2}}(I_1 I_2 - i X_1 X_2), \\
\XXp_{90}&\equiv&(X_{-90}\otimes X_{-90}) \XX_{90}.
\end{eqnarray}
The relationship between the $\XX_{90}$, $\XXp_{90}$ and controlled not ({\sc cnot}) gates is shown in Figure~\ref{xxcn} in circuit form.
$\XX_{90}$ and $\XXp_{90}$ have the useful property that the parity-encoded versions of these gates can be achieved by applying just one copy of the unencoded gate to a pair of physical qubits (where one physical qubit is selected from each of the encoded input blocks).

\begin{figure}
\begin{center}
\includegraphics[width=8cm]{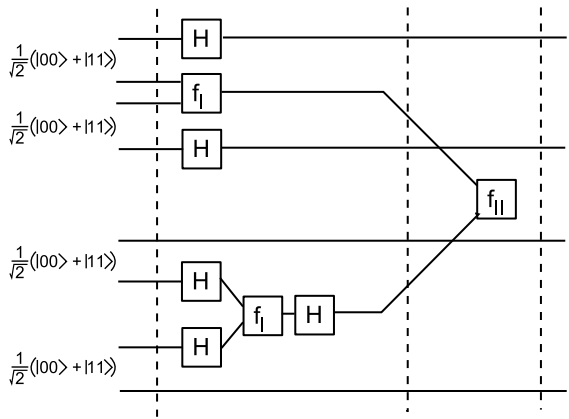}
\caption{The circuit used to create the resource state $\ket{R_{\XX}}$ used in implementing the $\XXp_{90}$ gate. The circuit can be modified by treating the top and bottom qubits as length-2 parity encoded qubits in order to improve the nondeterministic behaviour of the gate.}
\label{xxres}
\end{center}
\end{figure}

An unencoded $\XXp_{90}$ can be achieved nondeterministically as follows. First, a four-qubit resource state is created using the circuit shown in Figure~\ref{xxres}. If the three fusion gates are successful, the resulting state is
\begin{equation}
\ket{R_{\XX}}=\frac{1}{2}[\ket{++}(\ket{00}+\ket{11})+\ket{--}(\ket{01}+\ket{10})].
\end{equation}
Next, two $\fII$ gates are applied between the input qubits and the resource state in the following manner: a $\fII$ gate is applied between a qubit of the first input and the first qubit of the resource state, and a $\fII$ gate is applied between a qubit of the second input and the fourth qubit of the resource state. If both are successful, the remaining two qubits will be in the state $\XXp_{90}|\psi\rangle$, where $|\psi\rangle$ was the input state (subject to possible known Pauli corrections).

For parity-encoded inputs the procedure is almost identical, except that one has more opportunities to attempt the $\fII$ gates between the inputs and the resource state. If both of the $\fII$ gates fail, the corresponding encoded input qubits will be reduced in size by one and the gate can be re-attempted. Note that if one of the $\fII$ gates fails and the other succeeds, then one of the input states will be reduced in size by one but the other input effectively retains its size, if we consider one of the remaining qubits in the resource state to now belong to the particular encoded input that corresponds to the fusion gate that succeeded.

The procedure described above provides an encoded $XX'_{90}$ gate which on average decreases the size of its inputs by one. A simple modification can be made to the resource-state generation circuit so that the resulting gate will instead on average preserve the size of its inputs. The modification involves treating the first and fourth qubits in Figure~\ref{xxres} as length-2 parity qubits instead of physical qubits, meaning that the resource state is now a state of 6 physical qubits. We will make use of this version of the $\XXp_{90}$ gate in the remainder of the paper.

\section{Analysis of Errors}\label{secerrmod}

\subsection{Error Modelling}

The theory of error correction in general requires that some assumptions be made concerning the
nature of errors that may occur in a system. The set of possible errors that are considered, and
their probability, form an error model on which the conclusions of a theory are based. In developing
error correction methods, the aim is to choose error models which closely match the physical
reality. Typically, this involves focussing on the most common types of error found in the
corresponding experimental systems. However, it is not always possible to make the error model a
detailed fit to the requirements of a particular system, especially when the technology is still in
development.

In optical quantum computing systems with single photons, the greatest source of error is normally photon loss. This
can occur in many different ways, including tunnelling, detection failure, imperfect coupling and
unreliable sources. Another important type of error in optical systems is dephasing. It is expected that dephasing errors will typically be less frequent than loss errors in future optical computer components, but still likely to have a significant effect in any large-scale quantum circuit. These errors can also be caused by imperfect coupling, or indeed any misalignment or flaw in the optical elements that can affect the polarization of the photon.

We use a simple error model for the noise on physical qubits. Each gate operation is divided into timesteps,
with a single timestep being roughly the time required to make a measurement or set of measurements
and perform feedforward based on the results. Each physical qubit is considered to experience loss
at a rate of $\gamma$ per qubit per timestep, and Pauli errors at a rate of $\eta$ per qubit per
timestep. The Pauli error is selected randomly from the set $\{X$, $Y$, $Z\}$ with equal
probability. This corresponds to a depolarization error of rate $\frac{4}{3}\eta$, and to a marginal probability of a bit-flip of
$\frac{2}{3}\eta$. Depolarization errors are a generic way of representing the effects of dephasing noise, as well as other errors which act locally on each qubit and do not cause leakage from the qubit state space.

\begin{figure*}
\begin{center}
\includegraphics[width=17.5cm]{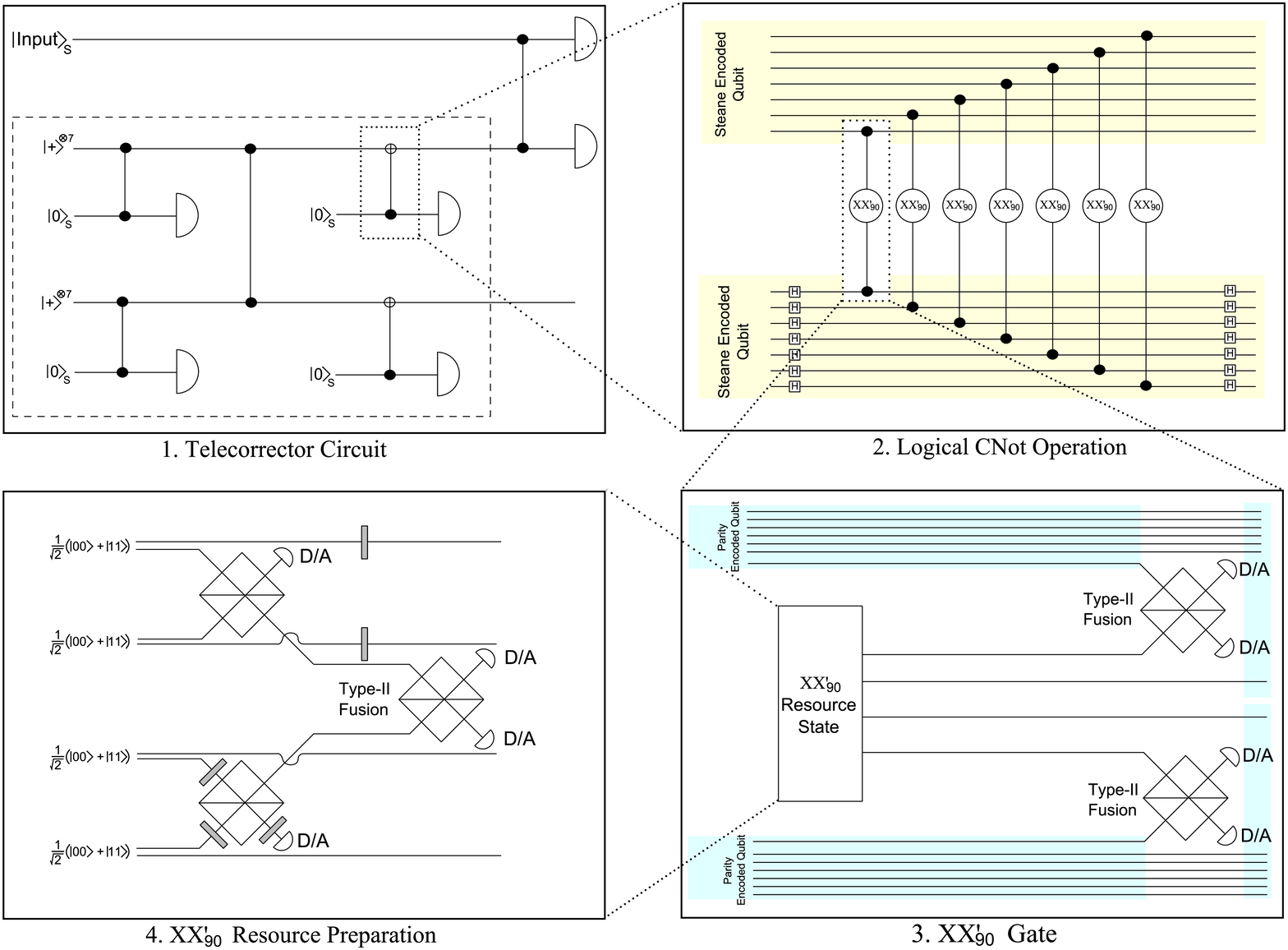}
\caption{The teleported error-corrector ({\em telecorrector}) circuit shown at different scales. This diagram demonstrates the manner
in which logical operations are broken down into a series of operations at the lower layers of
encoding. Box 1 shows the circuit for the production of the resource required for one round of
telecorrection. Box 2 demonstrates how a {\sc cnot} gate on the Steane-encoded qubits can be
performed using 7 $\XXp_{90}$ gates and 14 Hadamards at the parity encoding layer. Box 3 shows
the process of using resource preparation plus teleportation to implement an $\XXp_{90}$ gate, and
Box 4 provides the circuit used to prepare the required resource.} \label{zoomfig}
\end{center}
\end{figure*}

The physical error model described above is used as the basis for estimating the effective error rates on encoded qubits at higher levels of encoding.  The relationship between the levels of encoding are summarised in Figure~\ref{zoomfig}. At each encoding level, the aim is to estimate the rates of two different error types: {\em located errors}, which are those errors which are heralded (in a way analogous to the failure of fusion gates at the physical encoding level), and {\em unlocated errors}, which are Pauli errors that are not directly heralded and must be found indirectly via syndrome measurement. The remainder of this section is devoted to deriving expressions approximating the effective error rates of the various parity-encoded operations, i.e., operations at the first level of encoding above the physical level. Then in the following section, these expressions are used as the error model for simulations of concatenated fault-tolerant teleported error correction ({\em telecorrection}), using Steane and Golay coding.


In deriving expressions for the effective error rates at level 1, we use approximations which are correct only to first order in the physical error rate. In particular, we find the probability that a qubit remains error-free after multiple timesteps by taking the product of the individual probabilities of an error not occurring for each timestep. In general, this is correct only to first order, since in reality multiple errors of the same type occurring on the same qubit can cancel. For bit- or sign-flip errors of the type included in our error model, an even number of errors has no overall effect on the qubit. However, since we are only considering very small values of physical error rate, and since each parity-encoded gate consists of only a relatively small number of optical components, these higher order error terms are insignificant.

It is important to note that this model assigns the same error probability to a timestep regardless
of whether a qubit was involved in operations during that time. Hence this model takes into account
errors that arise while a qubit is kept in memory, and assumes that such error rates are similar to
those for a qubit actively involved in computation. This aspect of the model may be unduly
pessimistic, but we have chosen to consider the worst case in this regard.

Having defined the error model, it is also necessary to consider how these errors propagate through
the elements of an optical system. First, it should be noted that measurement in the $Z$ basis
renders any $X$ errors on that qubit irrelevant (likewise for $Z$ errors and $X$-basis measurements). Measurement also serves to locate
loss errors. Due to the regular measurements that occur in our protocol (either directly or via a fusion gate), it
can be seen that loss errors will always be quickly transformed into located errors.

The properties of parity state encoding are such that any $X$ errors on physical qubits immediately become $X$ errors on the
logical, parity-encoded qubit. Hence the probability of a logical $X$ error will depend simply upon
the rate of such errors at the physical level, the number of qubits and the duration of the
computation. $Z$ errors on individual photonic qubits do not automatically become logical, however if
a photonic qubit on which a $Z$ error has occurred is used as an input for a fusion gate, the error
is then applied to all the qubits forming the parity state of which the input qubit was a member. We will use $\eta_X = \eta_Z = \frac{2}{3}\eta$ to denote the marginal error probabilities of $X$ and $Z$ errors (where $Y$ is considered to be ``both an $X$ and a $Z$ error''). This notation allows one at a glance to see which physical error type contributes to a particular logical error rate in the expressions of the following subsections.


\subsection{Size of the Parity Code}

\begin{figure}
\begin{center}
\includegraphics[width=8.5cm]{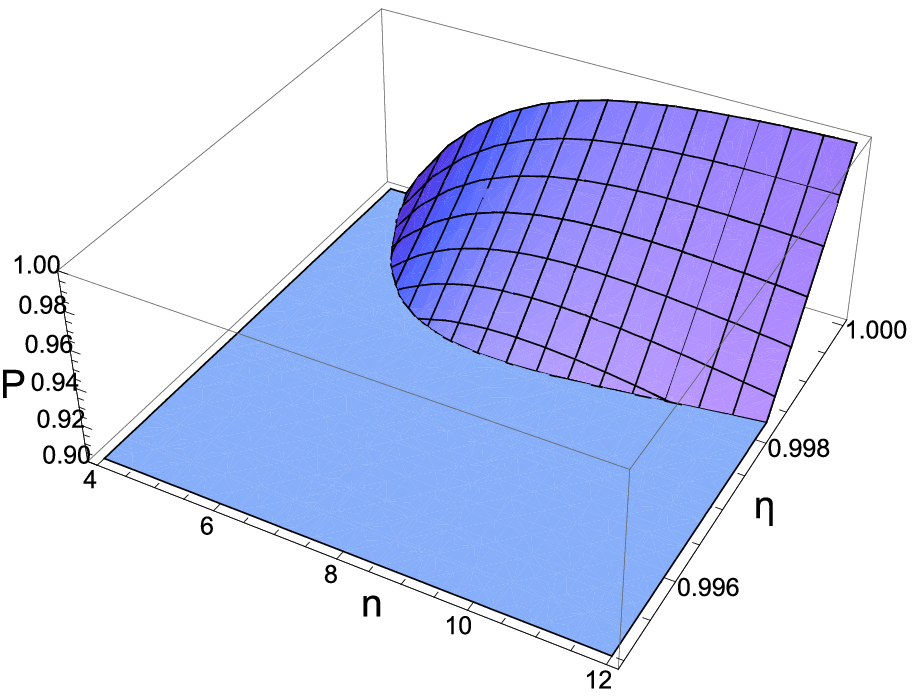}
\caption{The probability of success for the $\XX_{90}$ gate as a function of the photon loss rate ($1-\eta$) and code size $n$.} \label{tradoff}
\end{center}
\end{figure}

There is a trade-off in the located error rate for the non-deterministic gates between the errors due to gate failure and those due to loss
(Figure~\ref{tradoff}). This occurs as each additional qubit in the encoding increases the possibility of a loss occurring. However, a certain
level of encoding is necessary in order to reduce the probability of gate failure. We can optimise the size of the parity states by
calculating the located error rates for the non-deterministic operations examining how these errors vary with code size. Fortunately, the
optimal parity code size to minimize the trade-off is found to be similar for both non-deterministic operations, with a seven qubit code
proving to offer the best threshold. All error rate calculations will therefore assume a code of this size.

\subsection{Source Production}

We assume that copies of the state $\Ket{0}{7}$ are prepared as needed by massively parallel production in order to have them
ready when required by the circuit. We begin with Bell pairs in the state
$(\ket{00}+\ket{11})/\sqrt{2}$, and link these by means of type-I and -II fusion gates. For a
parity state of size 7, 7 Bell pairs and 6 fusion gates are required to build the resource. As each
gate has a 50\% failure rate, parallel production of each resource requires an average of $7\times
2^6$ Bell pairs, and takes 3 timesteps.  The resulting state has an unlocated $X$ error rate of $1-(1-\eta_X)^{41}$ and a located error rate of $1-(1-\gamma)^{21}$. (There are 41 locations in the optical circuit which can contribute errors to the output state. For 20 of these locations, photon loss will be immediately heralded, and so we post-select on no loss occurring at those locations).


\subsection{$Z_{90}$ Error Rates}

This gate occurs in two steps: the first step is the attempt to fuse the encoded state with a resource state, and the second step is measuring
the remaining component qubits from the original state once a successful fusion has been performed. The probability of a failure at the
first step is:
\begin{equation}
1-\sum_{j=1}^{7}2^{-j}(1-\gamma)^{\frac{1}{2}j(1+j)}
\end{equation}
which combines the possibility of a loss during fusion with the fusion failure rate to get the total located error rate during the fusion
attempts. To cover the possibility of loss occurring on any of the other component qubits, we include the factor
\begin{equation}
(1-\gamma)^{3j+(1+j)(7-j)}
\end{equation}
in the sum, which depends on the number of qubits remaining and the time they have been in memory.

Hence the $Z_{90}$ gate has a combined located error rate per parity qubit of
\begin{equation}
P_{LE}=1-\sum_{j=1}^{7}2^{-j}(1-\gamma)^{\frac{1}{2}j(1+j)}(1-\gamma)^{3j+(1+j)(7-j)}.
\end{equation}

For unlocated errors, the average rate can be found by combining the error rates for memory and for producing an ancillary parity state,
due to the re-encoding process used to implement the gate. The average time required to implement this gate would be 2 timesteps. However,
for all error types, it is assumed that the average time spent is 4 timesteps (this corresponds to the average time required for the slowest gate, $\XXp_{90}$). This is
done so that all encoded operation types can be treated as taking an equal amount of time. By counting the number of locations that may contribute to logical errors on the output, we obtain overall effective rates of unlocated $X$ and $Z$ errors of $1-(1-\eta_X)^{69}$ and $1-(1-\eta_Z)^{10}$ respectively, for the $Z_{90}$ gate.


\subsection{$\XX_{90}$ Error Rates}

The propagation of errors through this gate is fairly simple: a $Z$-error on either input qubit will
cause an $X$-error on the opposite qubit. $X$-errors on an input qubit propagate to the output without having an effect on the
other qubit.

As described in Section~\ref{subsec:xx90}, the gate is achieved by performing fusion gates between qubits from the encoded input state and a 6-qubit resource state.
As we are interested in
an error rate per parity qubit, we model the success or failure of the fusion gates applied to an input parity qubit using a random walk on a semi-infinite
1-dimensional lattice with 1 absorbing boundary at $-n$ \cite{redn01}. Here the lattice represents the size of the parity state, with failure occurring either due to a loss or when all component qubits are measured due to repeated teleportation failures.

The number of paths that reach the boundary at time $t$ is
\begin{eqnarray}
&N(n,t)=\frac{t!}{(\frac{t+n}{2}!)(\frac{t-n}{2}!)} &\text{for $(t-n)$ mod }2=0\\
&N(n,t)=0 &\text{for $(t-n)$ mod }2=1 \nonumber
\end{eqnarray}
where $n$ is the initial size of the parity code. To incorporate the possibility of a successful gate
operation, we include a scaling factor in the number of paths:
\begin{equation}
N_{\mathrm{scaled}}(n,t)=2^{-\frac{t-n}{2}}N(n,t).
\end{equation}
The number of first passage paths for this type of walk is
\begin{equation}
F(n,t)=\frac{n}{t}N_{\mathrm{scaled}}(n,t).
\end{equation}
Therefore, total probability of success per parity qubit in the absence of loss is:
\begin{equation}
P_S=1-\sum_{t=n}^{\infty}\frac{F(n,t)}{2^t}.
\end{equation}
For a parity qubit of size 7, this evaluates to $P_S=0.9763$.

As the average size of the state and the rate of loss per timestep is constant, the average loss
per parity qubit during this operation can be simplified to:
\begin{equation}
P_L=1-(1-\gamma)^{t(n+3)}.
\end{equation}
The average time taken for this operation is 4 timesteps. Thus, the approximate located error rate per parity-encoded input is $1-0.9763(1-\gamma)^{40}$. The unlocated $X$ and $Z$ error rates, obtained by counting error locations in the average-time case, are $1-(1-\eta_X)^{28}$ and $1-(1-\eta_Z)^4$ respectively.


\subsection{Memory and Measurement}

The memory or identity operation on an encoded qubit simply entails keeping the constituent physical qubits in memory while other operations
are performed. We treat the encoded memory operation as taking 4 time steps, equal to the average time taken to perform the slowest encoded gate, $\XXp_{90}$. This yields approximate rates of $X$, $Z$, and located errors of $1-(1-\eta_X)^{28}$ , $1-(1-\eta_Z)^4$, and $1-(1-\gamma)^{28}$ respectively.

An encoded computational-basis measurement involves measuring each physical qubit in the computational basis and finding the parity of the measurement results. Thus, the rates of located and unlocated errors in the measurement outcome are $1-(1-\gamma)^7$ and $1-(1-\eta_X)^7$ respectively.


\section{Fault Tolerance Threshold}\label{secthresh}

\begin{figure}
\begin{center}
\includegraphics[width=8.1cm, trim=6mm 6mm 5mm 0]{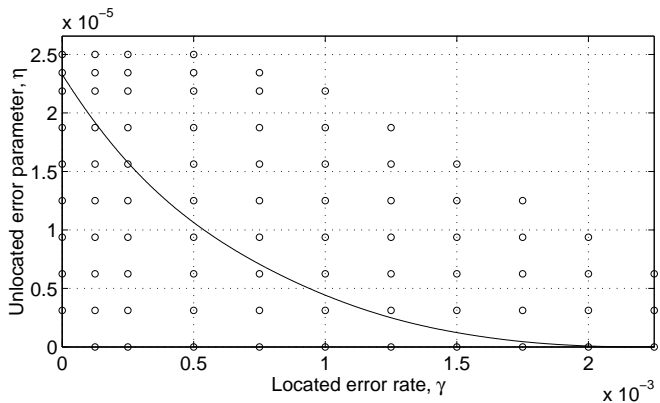}
\caption{The threshold curve for optical parity-state computing using the $7$-qubit Steane code. The region below the solid curve represents the set of error rates which can be tolerated by the scheme.} \label{stegraph}
\end{center}
\end{figure}

\begin{figure}
\begin{center}
\includegraphics[width=8.1cm, trim=6mm 6mm 5mm 0]{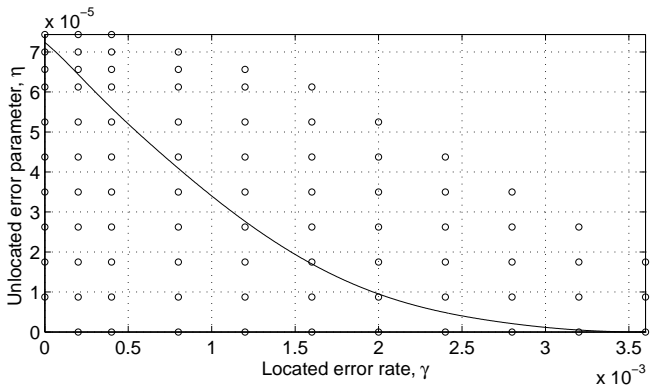}
\caption{The threshold curve for optical parity-state computing using the $23$-qubit Golay code.} \label{golgraph}
\end{center}
\end{figure}

Having analysed the emergence of logical errors at the parity code layer, we used numerical
simulation to calculate the error rates at higher levels of concatenation, and from this, the value of the noise threshold for optical parity-state quantum computation.

The simulations were performed using similar techniques to those described in Sec V.D of \cite{daw06}. In particular, we simulate one level of the telecorrector protocol, for both the $7$-qubit (Steane) and $23$-qubit (Golay) codes. Our simulation differs from \cite{daw06} in the noise model and gate set used. The telecorrector circuit in \cite{daw06} uses the following gate set: preparation of $|0\rangle$ and $|+\rangle$ states, {\sc cnot}, {\sc cphase}, and $X$-basis measurement. We converted this circuit to our gate set in the following way. First the circuit was expressed solely in terms of {\sc cphase} gates, Hadamard gates, $|+\rangle$ creation and $X$-basis measurements, by making appropriate substitutions of each {\sc cnot} gate, $|0\rangle$ creation, and computational basis measurements in the circuit, and simplifying the resulting circuit where possible. Then a simple change of basis $|0\rangle\leftrightarrow|+\rangle$, $|1\rangle \leftrightarrow |-\rangle$ yielded a circuit in our gate set ($\XXp_{90}$ gates, Hadamard gates, $|0\rangle$ creation, and computational-basis measurement).

We carried out a series of Monte Carlo simulations for a range of values of the physical error rates $(\gamma,\eta)$, in each case measuring the resulting rate of unlocated and located errors at the next highest level of encoding. For a particular choice of the physical noise rates, Pauli errors (both unlocated and located) are introduced by each gate with a probability that is governed by the noise model derived in the previous section. In the case of unlocated errors, $X$ and $Z$ errors are introduced independently. For the $\XXp_{90}$ gate, errors are introduced independently on each of the two output qubits.

The results of the simulations were used to characterise the mapping of error rates from the physical level to the second level of encoding (i.e., the parity encoding plus one level of telecorrection encoding) by way of a polynomial fit to the measured values. Our characterization of the mapping for all levels of encoding above this was obtained by simply using the particular polynomial that was obtained in \cite{daw06}. Thus, by applying the appropriate sequence of polynomials to a particular setting $(\gamma,\eta)$, we are able to estimate effective error rates at any level of encoding.

The {\em threshold curve} is defined to be the curve in the $\gamma-\eta$ plane below which effective error rates tend to zero for many levels of encoding.
Our calculated threshold curves are shown in Figures
\ref{stegraph}-\ref{golgraph} (the small circles show the values of $(\gamma,\eta)$ for which a Monte Carlo simulation was run).
Like previous schemes for optical quantum computing, the results demonstrate a trade-off between a tolerance of photon loss and depolarization errors. It is worth noting that the code is always required to deal with
some probability of located errors due to the non-deterministic gates and the finite size of the
underlying parity encoding. These errors are taken into account when calculating the threshold for
the loss and depolarizing rates plotted here.

\section{Resources}\label{secres}

\begin{figure}
\begin{center}
\includegraphics[width=7cm]{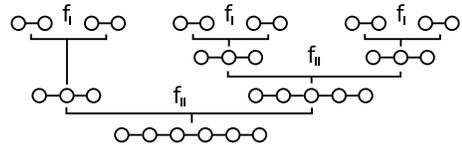}
\caption{Process for generating a 6 qubit parity state, beginning with 6 Bell states and performing
two rounds of fusion gates.} \label{resgen}
\end{center}
\end{figure}

For a useful comparison with other schemes for error-correcting quantum computing, it is necessary
to also consider the resources that would be required to implement this scheme. As noted
previously, the method considered here for the creation of resources involves parallel production
of many copies of a resource to ensure it is available on demand. This is a simple approach that
leads to a higher cost in terms of entangled photon generation, and avoids more complicated
resource-saving techniques such as storing previously prepared resources and ``recycling'' entangled
states from unsuccessful attempts. Such techniques tend to require more intricate design and
photon-switching, as well as potentially introducing more errors due to the longer photon storage
time.

We calculate the resources required in terms of the number of Bell states used. This a useful unit to consider as it is a common starting point for building entangled states across many optical schemes. It is also handy for comparisons with current optical quantum computation experiments as most use the entangled output of optical parametric down-conversion systems as photon sources.

As no recycling is used, the resource production has a 50\% chance of failing for every fusion gate
performed, meaning that the number of parallel attempts required doubles with each fusion gate.
This exponential growth can be tolerated as it only occurs on a small scale -- in our case none of the
resource states need to be larger than 8 qubits.
To estimate the resources we calculate the average number of Bell
states required to produce the appropriate resources using this parallel production method and
implement a parity-encoded operation.

Much of this resource production simply involves the creation of ancillary parity states, which is
done by linking the initial Bell states together using type-I and -II fusion gates (Figure~\ref{resgen}). State preparation at the parity layer requires a 7-qubit parity state, which would
take an average of 448 Bell pairs to produce. Re-encoding, and hence the $Z_{90}$ operation, would
require an 8-qubit parity state for each attempt, and an average of two attempts to implement the
operation. Hence we estimate the resources for the $Z_{90}$ operation at 2048 Bell pairs on
average. The circuit for the preparation of the $\XXp_{90}$ resource was shown previously in Figure~\ref{xxres}. This resource requires an average of 128 Bell pairs to produce.

As a basis for comparison, we will consider the requirements for producing the telecorrector state needed for one round of correction.
Using the gate costs described above, it can be calculated that the state requires approximately 177670 Bell pairs to generate. In Table~\ref{restable} resource usage and threshold values are compared between parity-state and cluster-state schemes for optical quantum computing. Due to its greater difference from the other two schemes, we did not include in the table a scheme for quantum computing using ``cat'' states \cite{lund08}. Cat states are in general more difficult to generate than Bell states, so this makes a direct comparison of resource usage difficult. We note that on average $10^3$ cat states are needed to create a telecorrector state in that scheme. The loss threshold for the cat states is $2\times10^{-4}$, and a threshold for depolarization was not specified in that work.

Thus, our scheme for fault-tolerant parity-state quantum computing gives a resource usage figure three orders of magnitude smaller than that for an equivalent cluster state circuit \cite{daw06} (and, in some sense, three orders of
magnitude larger than the requirements for the coherent state version \cite{lund08}). However threshold is poorer than that of the cluster state protocol, but better (with respect to loss) than the scheme using cat states.

\begin{table}
\begin{tabular}{|c|c|c|c|}
\hline
&Loss&Depolarization&\\
Scheme&Threshold&Threshold&Resources\\
\hline
Cluster states&$4\times10^{-3}$&$8\times10^{-5}$&$1.3\times10^8$
\footnote{Referece \cite{daw06} quotes an incorrect resource usage number, the one cited here is the corrected value.}\\
Parity states&$2\times10^{-3}$&$2.4\times10^{-5}$&$1.8\times10^5$\\
\hline
\end{tabular}
\caption{A comparison of thresholds and resources for linear optics quantum computing error correction schemes using a 7-qubit Steane code. Here the resources are those required for the first level of telecorrection. } \label{restable}
\end{table}

\section{Conclusions}

We have shown that an error-correcting system based on parity encoding falls
in-between other schemes in both threshold and resource requirements. The parity scheme has a
higher error threshold than that found for coherent states, but also significantly larger resource
requirements. Conversely, it is two orders of magnitude cheaper in resources than a cluster-state
implementation, but also has lower thresholds for both located and unlocated error rates. Together,
these results suggest a necessary trade-off between resources and achievable threshold, which indicates that the preferred encoding method in any experimental attempt to demonstrate optical
quantum error correction will depend on the capabilities and limitations of the physical system and
its components.

It is worth noting the general principle demonstrated here, that concatenation can be used to tailor error
correction to suit the relative error rates. In this case, some parity encoding is used to reduce
the rate of failures due to non-deterministic gates to a level at which the remaining errors can be
handled by the general error correction code. However, the same principle could be applied to other
errors. For example, a system with a very high loss rate in comparison with other errors could use
concatenation with a dedicated loss-correction code \cite{ral05} to allow a higher threshold for loss at the
cost of lower thresholds for other errors.

\bibliography{references}

\end{document}